\documentstyle[12pt]{article}

\catcode`\@=11
\newdimen\@rotdimen
\newbox\@rotbox

\def\@vspec#1{\special{ps:#1}}
\def\@rotstart#1{\@vspec{gsave currentpoint currentpoint translate
   #1 neg exch neg exch translate}}
\def\@rotfinish{\@vspec{currentpoint grestore moveto}}
\def\@rotr#1{\@rotdimen=\ht#1\advance\@rotdimen by\dp#1%
   \hbox to\@rotdimen{\hskip\ht#1\vbox to\wd#1{\@rotstart{90 rotate}%
   \box#1\vss}\hss}\@rotfinish}
\def\@rotl#1{\@rotdimen=\ht#1\advance\@rotdimen by\dp#1%
   \hbox to\@rotdimen{\vbox to\wd#1{\vskip\wd#1\@rotstart{270 rotate}%
   \box#1\vss}\hss}\@rotfinish}%
\def\@rotu#1{\@rotdimen=\ht#1\advance\@rotdimen by\dp#1%
   \hbox to\wd#1{\hskip\wd#1\vbox to\@rotdimen{\vskip\@rotdimen
   \@rotstart{-1 dup scale}\box#1\vss}\hss}\@rotfinish}%
\def\@rotf#1{\hbox to\wd#1{\hskip\wd#1\@rotstart{-1 1 scale}%
   \box#1\hss}\@rotfinish}%
\def\rotate{\@ifnextchar[{\@rotate}{\@rotate[l]}}
\def\@rotate[#1]#2{\setbox\@rotbox=\hbox{#2}\@nameuse{@rot#1}\@rotbox}

\catcode`\@=12

\def\NPB#1#2#3{Nucl. Phys. B {\bf#1} (19#2) #3}
\def\PLB#1#2#3{Phys. Lett. B {\bf#1} (19#2) #3}

\def\PRD#1#2#3{Phys. Rev. D {\bf#1} (19#2) #3}
\def\PRL#1#2#3{Phys. Rev. Lett. {\bf#1} (19#2) #3}

\def\ov{\overline}
\def\s2{\frac{1}{\sqrt2}}

\def\beq{\begin{equation}}
\def\eeq{\end{equation}}
\def\beqa{\begin{eqnarray}}
\def\eeqa{\end{eqnarray}}
\def\IZ{\relax\ifmmode\hbox{\ss Z\kern-.4em Z}\else{\ss Z\kern-.4em Z}\fi}
\def\IP{\relax{\rm I\kern-.18em P}}
\def\IC{\relax\hbox{\kern.25em$\inbar\kern-.3em{\rm C}$}}
\def\hm{\relax{n_H}}
\def\vm{\relax{n_V}}
\def\cp#1{\relax\ifmmode {\IP\kern-2pt{}_{#1}}\else $\IP\kern-2pt{}_{#1}$\fi}

\topmargin
-2cm
\textwidth
15cm
\textheight
24cm
\oddsidemargin
1cm
\evensidemargin
1cm
\begin{document}
\date{}
\title{ Chains of N=2, D=4  heterotic/type II duals}

\author{G. Aldazabal \thanks{Permanent Institutions:
CNEA, Centro At\'omico
Bariloche, 8400 S.C. de Bariloche, and CONICET, Argentina.} $^1$,
A. Font \thanks{On sabbatical leave from
Departamento de F\'{\i}sica, Facultad de Ciencias,
Universidad Central de Venezuela. Work supported in part by the
N.S.F. grant PHY9511632 and the Robert A. Welch Foundation.} $^2$,
L.E. Ib\'a\~nez$^{1,3}$ and F. Quevedo$^4$ \\[4mm]
$^1$ \normalsize Departamento de F\'{\i}sica Te\'orica, \\[-3mm]
\normalsize Universidad Aut\'onoma de Madrid, \\[-3mm]
\normalsize Cantoblanco, 28049 Madrid, Spain.
\\[3mm]
 $^2$\normalsize Theory Group, Department of Physics,\\[-3mm]
\normalsize The University of Texas,\\[-3mm]
\normalsize Austin, TX 78712, USA.
\\[3mm]
$^3$\normalsize Department of Physics and Astronomy, \\[-3mm]
\normalsize Rutgers  University, \\[-3mm]
\normalsize Piscataway, NJ 08855-0849, USA.
\\[3mm]
 $^4$ \normalsize Theory Division, CERN, 1211 Geneva 23,
Switzerland.
}
\maketitle
\vspace{-5.4in}
\rightline{CERN-TH/95-270, FTUAM-95/38}
\rightline{RU-95-65, UTTG-17-95}
\rightline{hep--th/9510093}
\vspace{5.0in}
\begin{abstract}
We report on a search for $N=2$ heterotic
strings that are dual candidates of type II
compactifications
on Calabi-Yau threefolds described as $K3$ fibrations.
We find many new heterotic duals  by
using standard orbifold techniques.
The associated type II compactifications
fall into chains in which the proposed duals
are heterotic compactifications
related one another by a sequential Higgs mechanism.
This breaking in the heterotic side
typically involves the sequence
$SU(4)\rightarrow SU(3)\rightarrow $ $SU(2)\rightarrow 0$,
while in the type II side the weights of the complex hypersurfaces
and the structure of the $K3$ quotient singularities
also follow specific patterns.
Some qualitative features of the relationship
between each model and its dual can be understood by
fiber-wise application of string-string duality.

\end{abstract}
\maketitle

\newpage

\section{Introduction}

In the last few months, evidence has been found in favor
of a strong-weak coupling duality between type II strings
compactified on certain Calabi-Yau (CY) manifolds and certain
$N=2$ heterotic strings in four dimensions \cite{kv, fhsv, klm,
klt, vw, agnt, kklmv}.
Highly non-trivial perturbative and non-perturbative checks
have been performed for a few
pairs of duals in which the type IIA strings are compactified
on CYs with small number of K\"ahler deformations.

In spite of this recent progress, at the moment there is no
general construction that produces the heterotic dual of a given type II
compactification. In fact,
the general idea \cite{kv} is not that any model has a
dual description but rather that there are
some string models that admit both type II and heterotic dual
realizations whereas there will be models that admit only one
or the other (or none). Of course, finding examples with
both type II and heterotic dual interpretations is extremely
interesting because using both descriptions simultaneously allows to extract
non-perturbative information about the relevant $N=2$ theory considered.
In this context, it has been pointed out \cite{klm} that
CY compactifications corresponding to $K3$ fibrations
seem to play an important r\^ole in heterotic/type II duality.

Our understanding of $D=4$, $N=2$ type II/heterotic duality is also
quite incomplete regarding the issue of how the spaces of models are
connected. Indeed, it has been proposed \cite{kv}
that stringy gauge symmetry enhancement
may provide a way of continuously connecting
many (or all) $N=2$ heterotic vacua. It would be interesting to
see  explicitly in more detail how such a `web' of
heterotic vacua is actually formed. Obviously, there is also the
question of how this web in the heterotic side maps into the
type II side, yielding compactifications connected somehow.
In fact, it is known that CY spaces are connected along paths
where conifold singularities develop \cite{coni}.
In the type IIB theory
these singularities appear in the moduli space of vector
multiplets and the transition to a different CY can be explained
in terms of blackhole condensation \cite{ast, gms}.
We would like to know how these transitions translate into
the heterotic side. Moreover, we would also need to understand
the problem of singularities in the moduli space of hypermultiplets
\cite{bbs}.

It seems clear that in order to address some of the above
issues,
as well as to extend this duality to the $N=1$ case,
a better understanding of the space of $N=2$
heterotic compactifications is needed.
In comparison, the CY threefolds in the type II side are
much better known and, due to its application
to $N=1$ heterotic compactifications, there exist
long lists of models with different topological data.
Such a systematic study in
the heterotic $N=2$ case is lacking.

In the present note we begin a systematic exploration of $N=2$
heterotic models. Most of our examples are obtained by
compactifying the heterotic string on symmetric orbifolds
$T^4/Z_M\times T^2$ and simultaneously
embedding the $Z_M$ symmetry in the gauge degrees of freedom.
Besides the fact that the orbifold conformal field theory
and partition function are well known, there is perhaps another
naive motivation to use this kind of compactification towards
constructing pairs of heterotic/type II duals. This
duality is supposed to have its roots on an underlying
string-string duality \cite{stst}
between the type IIA string compactified on
$K3$ and the heterotic string compactified on a 4-torus.
Considering $K3$ fibrations on the type II side and using
string-string duality fiber-wise, it has been argued \cite{vw}
that on the heterotic side the $K3$ fibers should be replaced
by $T^4$ fibers.

Starting with an specific $N=2$ heterotic parent we derive
chains of descendant models obtained by appropriate Higgsing,
both using hypermultiplets and vector multiplets. In  many
examples we find that the last four elements of the chains,
typically  involving the sequential gauge breaking
$SU(4)\rightarrow SU(3)\rightarrow SU(2)\rightarrow  0$,
have candidate type II duals that appear in the lists of
$K3$ fibrations in ref.~\cite{klm} . Moreover, the weights of the
corresponding weighted projective spaces
follow specific sequential patterns. These patterns
also reflect in a certain structure of the quotient singularities
of the $K3$ fiber.
For example, to the chain of
heterotic breakings $SU(3)\rightarrow SU(2)\rightarrow 0$
there corresponds $K3$ singularities of
type $A_i\rightarrow A_{i-1}\rightarrow A_{i-2}$.
We also find that many of the $K3$ fibrations listed
in ref.~\cite{klm} can be put into similar sequences, although
heterotic duals for all of them are not yet available.
These sequences of CY models presumably correspond to manifolds
connected through some sort of transition
in the type II language.

This note is organized as follows. In section 2 we briefly review
orbifold compactifications. In section 3 we construct chains of
heterotic models and conjecture their type II duals. In section 4
we discuss the structure of the chains.
Finally, in section 5 we present our conclusions and
outlook.

\section{Constructing $N=2$ heterotic models}

There are many possibilities available to build
$N=2$, $D=4$ heterotic models with different gauge groups.
They fall essentially into two classes: left-right symmetric and
asymmetric. Examples of the first class are obtained
by compactifying on $K3\times T^2$ and
simultaneously embedding the spin connection into the
gauge degrees of freedom in a modular invariant manner.
There are many possible ways to construct the $K3$ and also many
possible modular invariant gauge embeddings.
An alternative to following the `Calabi-Yau'
approach of ref.~\cite{kv}, is to consider
exact conformal field theory (CFT) constructions.
During the past years several formalisms have been developed
to construct $N=1$ heterotic CFT's based on free or coset
theories (for a collection of some relevant papers
see ref.~\cite{schell}). Extending these techniques to $N=2$,
myriads of left-right symmetric models can be constructed for
which the CFT is known and therefore the couplings can
be explicitly computed.

A simple start is provided by symmetric toroidal
orbifold compactifications on $T^4/Z_M\times T^2$.
Acting on the (complex) bosonic transverse coordinates,
the $Z_M$ twist $\theta$ has eigenvalues $e^{2 \pi i\, v_a}$,
where $v_a$ are the components of $v=(0,0,\frac1M,-\frac1M)$.
Unbroken $N=2$ SUSY requires $M=2,3,4,6$ \cite{dhvw}. The embedding
of $\theta$ on the gauge degrees of freedom is usually
realized by a shift $V$ such that $MV$ belongs to the $E_8\times E_8$
or $Spin(32)/Z_2$ lattice.
This shift is restricted by the modular invariant
constraint $M\, (V^2-v^2)={\rm even}$. All possible embeddings for
each of the four allowed orbifolds can be easily found.

In the $E_8\times E_8$ case, we find 2 inequivalent embeddings
for $Z_2$, 5 for $Z_3$, 12 for $Z_4$ and 59 for $Z_6$,
leading to different patterns of $E_8\times E_8$ symmetry
breaking to rank-16 subgroups
\footnote
{The inequivalent $Z_4$ and $Z_6$ shifts for
a single $E_8$ have been classified in refs.~\cite{homy, kat}.
Those results can be used to find the combinations satisfying
modular invariance in the $N=2$ case.}.
A similar analysis can be carried out in the $Spin(32)/Z_2$ case.
Each of these models is only the starting point
for a big class of models generated by adding Wilson
lines in the form of further shifts in the gauge
lattice satisfying extra modular invariant constraints,
permutations of gauge factors, etc..
The possibility of enhanced symmetry groups,
at special points in the six-torus moduli
space, can also be considered. We will not analyze those
generalizations here. We will also restrict to left-right
symmetric twists.

To find the spectrum for each model, we can easily adapt the
analysis of the $N=1$ compactifications \cite{sierra} to the $N=2$
case. There are $M$ sectors twisted by $\theta^n, n=0,1,\cdots , M-1$.
Each particle state is created by a product of left and right
vertex operators  $L\otimes R$.
At a generic point in the six-torus moduli
space, the massless states follow from
\beq
\label{uno}
m_R^2=N_R+\frac{1}{2}\, (r+n\,v)^2
+E_n-\frac{1}{2} \quad ;\quad
m_L^2=N_L+\frac{1}{2}\, (P+n\,V)^2+E_n-1
\eeq
Here $r$ is an $SO(8)$ weight with $\sum_{i=1}^4 r_i=\rm odd$
and $P$ a  gauge lattice
vector with $\sum_{I=1}^{16} P^I= \rm even$.
$E_n$ is the twisted oscillator contribution to the
zero point energy and it is given by $E_n=n(M-n)/M^2$.

The multiplicity of states satisfying eq.~(\ref{uno}) in a
$\theta^n$ sector is given by:
\beq
\label{dos}
D(\theta^n)=\frac{1}{M}\sum_{m=0}^{M-1}\chi(\theta^n,\theta^m)\,
\Delta(n,m)
\eeq
where
\beq
\label{tres}
\Delta(n,m)=\exp\{2\pi\, i[(r+nv)\cdot mv-
(P+nV)\cdot  mV + \frac{1}{2}mn(V^2-v^2)+m\rho]\}
\eeq
In the above $\chi(\theta^n,\theta^m)$ is a numerical factor
that takes into account the fixed point degeneracy. More precisely,
$\chi(1,\theta^m)=1$, implying that for the
untwisted sector $D(1)$ only projects out the
toroidal states which are not invariant under the
twist. Otherwise, $ \chi(\theta^n,\theta^m)$ is the number of
simultaneous fixed points of $\theta^n$ and $\theta^m$. Recall that
$\theta$ acts on $T^4$ and always leaves fixed points,
 unlike the $N=1$ case where there are some
twists with fixed tori. The factor $\rho$ only appears in the case
of oscillator states ($N_L\neq 0$), the phase $e^{2\pi\, i\rho}$
indicates how the corresponding oscillator is rotated by $\theta$.

Let us now consider the spectrum in the different sectors.
In the untwisted sector, the right-moving massless states
have $N_R=0$ and $r^2=1$,
the left-moving states come either from $N_L=1$ or $P^2=2$.
The $L\otimes R$ invariant states are selected by the condition
$ P\cdot V - r\cdot v- \rho=\rm int$, with $\rho=0,1/M,-1/M$.
The left-moving oscillators from the spacetime coordinates combine
with the right-moving states with $r\cdot v=0$ to generate the
$N=2$ supergravity multiplet, including the graviphoton and the
vector-dilaton, giving rise to a model independent $U(1)^2$.
The same right-moving states combined with the left-moving
oscillators corresponding to the unrotated internal complex coordinate,
generate two extra $U(1)$ vector multiplets. This $U(1)^2$ symmetry
can be enhanced to a non-Abelian group for particular values
of the moduli of the unrotated 2-torus.
The model dependent non-Abelian vector multiplets are given by the
$r\cdot v=0$ states combined with the gauge lattice
vectors satisfying $P\cdot  V=\rm int$ and with gauge oscillators.
The states  with $P\cdot V=1/M + \rm int $ combine with
$r\cdot v=1/M$ to generate charged matter hypermultiplets.
The corresponding antiparticles come from $r\cdot  v =-1/M$.
Finally, the
moduli hypermultiplets come from the same right-moving states
combined with the oscillators corresponding to the two rotated
coordinates ($\rho=\pm 1/M$). These are two model independent
gauge singlets (they happen to be enhanced to four singlets for the
$Z_2$ orbifold.)

The twisted sectors contain only matter hypermultiplets.
For instance, the right-moving sector
of the $\theta$-twisted sector is always given by the two
vector  plus  two spinorial weights of $SO(8)$
satisfying $r^2=1,\, r\cdot v=-1/M$. They correspond
to the degrees of freedom of an $N=2$ hypermultiplet whose
gauge quantum numbers, coming from the left-moving
sector, depend on the model. The antiparticles of
an $n$-twisted sector arise in the $(M-n)$-twisted
sector. For sectors of order two ($n=M/2$), particles and
antiparticles are in the same sector, thus in
order not to overcount the states the corresponding
degeneracy is half of what is obtained in
equation (\ref{dos}).

In order to be more concrete, we now discuss the spectrum of
a particular model. We will consider  the
$Z_4$ orbifold with standard embedding $V=\frac14(1,-1,0,\cdots ,0)$
in one single $E_8$. The gauge group is given
by the roots $P$ satisfying $P\cdot  V=0$, breaking the group to
$E_7\times U(1)\times E_8\times U(1)^4$. The untwisted matter
is given by
 the roots satisfying $P\cdot  V=1/4$ leading to an
$N=2$ hypermultiplet transforming
under $E_7\times U(1)$ as a  $(56,1)$
\footnote{
The $U(1)$ charge is computed by the scalar product $(P+nV)\cdot Q$,
where $Q=(1,-1,0,\cdots ,0)\times (0,\cdots ,0)$ is the (unnormalized)
$U(1)$ generator.
All matter is singlet under the unbroken $E_8\times U(1)^4$ group.}.
Since for these states $P\cdot V - r\cdot  v=0$, they survive the
orbifold projection. To these states we have to add the
two model independent singlets $2\, (1,0)$ mentioned above.

The sector twisted by $\theta$ has several states
depending on the value of $N_L$.
For $N_L=0$ the massless states are given by the $E_8$ weights
satisfying $(P+V)^2=13/8$ that produce a $(56,-1/2)$
representation. The degeneracy factor is 4,
since the number of fixed points
of a $Z_4$ twist in  $T^4$ is 4, i.e.
$\chi(\theta,\theta^m)=4$. For $N_L=1/4$, the weights satisfy
$(P+V)^2=9/8$ and there are two left-moving oscillator states,
corresponding to the two rotated coordinates, with $\rho=-1/4$.
Again equation (\ref{dos}) gives an overall multiplicity number
of 4 leaving then a  total of 8 singlets
$(1,-3/2)$. For $N_L=3/4$ there are a total of $6$ oscillators
with $\rho=-3/4$ and $D(\theta)=4$ for the solution of $(P+V)^2=1/8$ ,
implying $24\, (1,1/2)$.

For the $\theta^2$ sector, the multiplicity factor  takes
the form $D(\theta^2)= \frac{1}{2}(4+\Delta(2,1))\, (1+\Delta(2,1)^2)$.
For $N_L=0$ there are $5\, (56,0)$ from solving $(P+2V)^2=3/2$
which has $\Delta(2,1)=1$.
For $N_L=1/2$ there are two solutions of $(P+2V)^2=1/2$ for which
$(P+2V)\cdot  V=\pm1/4$. Each combines with two pairs of
oscillators with
$\rho=\pm 1/4$, giving $\Delta(2,1)=\pm 1$ and $D=5,3$ and
altogether making 16 copies of $(1,1)+(1,-1)$.

The total matter spectrum is
\beqa
\label{cuatro}
\theta^0&:&(56,1)+2\, (1,0)\nonumber\\
\theta^1&:&4\, [(56,-\frac{1}{2})+2\, (1,-\frac{3}{2})+6\,
(1,\frac{1}{2})]\nonumber\\
\theta^2&:& 5\, (56,0)+16\, [(1,-1)+(1,1)]
\eeqa
Notice that in total there are $66$ singlets of $E_7$
as expected
from the $K3$ moduli space (since one is used to break $U(1)$).
Notice also that we have to be very careful  in
finding the number of gauge singlets (which are neglected in
many discussions) since they play a crucial
role in the identification of heterotic/type II
dual pairs as we will see next.

\section{Chains of heterotic duals}

In the construction of heterotic $N=2$ models, our main interest will be
in finding examples whose number of hypermultiplets ($\hm$) and vector
multiplets ($\vm$) matches the number of such multiplets in a type IIA
compactification on a CY threefold with Hodge numbers $b_{11}$ and
$b_{21}$. Since the dilaton lives in a vector multiplet in the heterotic
side, but in a hypermultiplet in the type IIA side, it must be that
$(\hm, \vm) = (b_{21}+1, b_{11}+1)$.
Furthermore, due to its moduli structure, it is expected
that the appropriate CY manifolds
for type II compactification should be understandable as $K3$ fibers on
$\cp1$ \cite{klm}. The conjectured underlying string-string duality in
six dimensions supports this interpretation \cite{vw}. Indeed, all
examples of heterotic/type II dual pairs analyzed up to now do correspond
to CYs that are $K3$ fibrations. Two lists of such manifolds were provided
in ref.~\cite{klm} . The first list includes 31 simple hypersurfaces
in weighted $\cp4$ in which the associated $K3$ fibers are a subset
of the 95 $K3$ transversal families classified in ref.~\cite{ponja}.
The second list gives 25 additional $K3$ fibrations which are
complete intersection CY spaces in weighted $\cp5$.
As a first exercise we will try to find $N=2$ heterotic
compactifications that match the spectra of type IIA
theories compactified on those CYs.

Let us now describe our strategy. We will start with an specific
heterotic $N=2$ orbifold and use the hypermultiplets
to break the  gauge symmetry by the Higgs mechanism step by step,
decreasing the rank of the group in one unit in each step
(more complicated possibilities will be mentioned below).
In principle, care must be taken not to
spoil the $N=2$ symmetry by giving vevs along non-flat directions.
However, that is specially easy in an $N=2$ theory.
Looking at it from the $N=1$ point of view, it is enough to impose
the usual $N=1$ D-flatness conditions and in addition an
F-flatness condition coming from the $N=1$ Yukawa coupling
between the adjoint chiral field inside the $N=2$ vector
multiplet and the chiral fields inside the hypermultiplets.
Both conditions together may be seen as D-flatness conditions
of the $N=2$ theory. Flat directions respecting $N=2$
do in general exist .
After each step of symmetry breaking
we give generic vevs to the adjoint scalars in the
(unbroken) vector multiplets.
This has two effects, namely
it gives masses to any hypermultiplet
charged with respect to the rank-reduced group, and it breaks
the rank-reduced gauge group to its maximal Abelian subgroup.
The final result will be a model with gauge group $U(1)^n$,  with
$2\leq n\leq  20$. As explained in the previous section, the lower
limit comes from the
multiplets containing the graviphoton and the dilaton,
whereas the upper limit is the
maximal rank achievable for generic values of the
$T^4/Z_M$ or $K3$  moduli.
Notice that the breaking of the gauge group down to its Abelian subgroup
is necessary to match the type II side whose (perturbative) gauge group
is just  $U(1)^{b_{11}+1}$. Furthermore, the hypermultiplets must be neutral
with respect to the $U(1)$s, which is guaranteed by our construction
in the heterotic side.

We have performed a systematic search of chains of $N=2$ models
following the above procedure and starting mostly with orbifold
compactifications.  We will spare the reader the details of all
these models and show the most relevant heterotic examples
found up to now in our search. We hope to report
more complete results in a future publication.
As we said above, we have obtained several
new heterotic models that match the lists of $K3$ fibrations in
ref.~\cite{klm}. One of the most interesting results is the existence of
five chains, each of four CY spaces, whose conjectured duals are given by
heterotic $N=2$ models in which a `cascade  gauge symmetry breaking'
takes place. Each such chain has very much the same structure.
For instance, the four models in the chain have a gauge group
(before adjoint Higgsing) of the form  $SU(m)\times G_r$,
with $m=4,3,2,0$. $G_r$ is an additional gauge factor
of rank  $r=12,10,8, 4$ and $11$ for each of the five chains
respectively.

We now describe each chain labelled by the value of $r$.

{\bf  1)  $r=12$  chain}

This chain may be obtained by appropriate Higgsing of the
$Z_2$ orbifold with standard embedding and gauge group
$E_7\times SU(2)\times E_8\times (U(1)^4)$.
The hypermultiplets, transforming only under $E_7\times SU(2)$, are
\beqa
\theta^0&:&(56,2)+4\, (1,1)\nonumber\\
\theta^1&:&8\, [(56,1)+ 4\, (1,2)]
\label{spz2}
\eeqa
Higgsing away the $SU(2)$ we are left with $65$ singlet
hypermultiplets. Now we give a vev to the adjoint Higgses inside
$E_7\times E_8$
and break it down to $U(1)^{15}$, while all $56$-plets get a mass.

We are left altogether with $65$ singlet hypermultiplets and
19 $U(1)$s, i.e. a model of type $(\hm,\vm)=(65,19)$
in the notation of \cite{kv}. In fact, this is nothing
but the first of a series of models
with $(\hm,\vm)=(65,19),\ (84,18),\ (101,17),\ (116,16)$ already constructed
by Kachru and Vafa. They are obtained by a
`cascade breaking'  $E_7 \rightarrow E_6
\rightarrow SO(10)\rightarrow  SU(5)$.  None of these models matches the
Hodge numbers given in the tables in ref.~\cite{klm}. However, the
interesting results are obtained by continuing the breaking
through $SU(5)\rightarrow SU(4)$ $\rightarrow SU(3) \rightarrow  SU(2)$
$\rightarrow 0$. In this case it can easily be checked
that a chain of models with
$(\hm,\vm)=\ (167,15),\ (230,14),\ (319,13),\ (492,12)$ are generated.
All of them have dual candidates in the lists in ref.~\cite{klm}.

{\bf  2)  $r=10$ chain }

The starting point is
one of the four possible non-standard  $E_8\times E_8$
embeddings of the $Z_3$ orbifold, with gauge shift
$V=\frac13(1,-1,0,\cdots ,0)\times \frac13(1,1,-2,0,\cdots ,0)$.
The gauge group turns out to be
$E_6\times SU(3)\times E_7\times U(1)\times (U(1)^4)$.
The massless hypermultiplet spectrum is easily found
along the lines discussed in the previous section. Explicitly,
\beqa
\label{cinco}
\theta^0&:&(27,3;1,0)+ (1,1;56,1)+(1,1;1,-2)+2\, (1,1;1,0)\nonumber\\
\theta^1&:&9\, [(27,1;1,\frac{2}{3})+ (1,{\overline 3};1,-\frac{4}{3})+2\, (1,
{\overline 3};1,\frac{2}{3})]
\eeqa
We now Higgs the group $E_7\times SU(3) \times U(1)$ as much as possible.
The two last factors can be Higgsed away completely whereas the $E_7$
can only be broken to a subgroup of rank 6 (e.g., $E_6$) since there is only
one $56$ available for Higgsing. Now, giving a generic vev to the adjoint
of this rank-6 group we are just left with an unbroken
$E_6\times U(1)^6\times (U(1)^4)$ group with
$12\, (27) +76\, (1)$ hypermultiplets.
We now proceed as in the previous chain by sequential Higgsing
$E_6\rightarrow SO(10)\rightarrow  $$SU(5)\rightarrow  ...$
$SU(3)\rightarrow SU(2)\rightarrow 0$.
In this way we obtain models with
$(\hm,\vm)=(76,16),\
(87,15),\ (96,14),\ (129,13),\ (168,12),\ (221,11),\ (322,10)$.
Again, the last four steps, corresponding
to the sequential breaking
$SU(5)\rightarrow SU(4)\rightarrow  \cdots \rightarrow 0$
have counterparts in the lists in ref.~\cite{klm}.

{\bf 3) $r=8$ chain }

We consider the $E_8\times E_8$
embeddings of the $Z_4$ orbifold with  gauge shift
$V=\frac14 (1,1,1,-3,0,\cdots ,0)\times \frac14(1,1,-2,0,\cdots ,0)$.
The gauge group in this example is
$SO(10)\times SU(4)\times E_6\times SU(2) \times U(1)\times (U(1)^4)$.
The massless hypermultiplets content is given by
\beqa
\label{seis}
\theta^0&:&(16,4;1,1,0)+(1,1;27,2,1)+(1,1;1,2,-3)+2\, (1,1;1,1,0)\nonumber\\
\theta^1&:&4\, [(16,1;1,1,\frac{3}{2})+(1,{\overline 4};1,2,-\frac{3}{2})
+2\, (1,{\overline 4};1,1,\frac{3}{2})]\nonumber\\
\theta^2&:& 5\, (10,1;1,2,0)+3\,  (1,6;1,2,0)
\eeqa
This model has already an $SU(4)$ group at the start,
so that a possibility would be to
Higgs away as far as possible the rest of the
gauge group and then start breaking
$SU(4)$ step by step. How far down can one break the rest of the group?
It is obvious that the $SO(10)$ group can be broken completely since there are
enough $16$-plets and $10$-plets to do the job.

On the other hand, we cannot Higgs
away completely the $E_6$ factor, since there are only $2\, (27)$s.
After examining the possible Higgsings, we conclude that the maximal
breaking is $E_6\times SU(2)\times U(1) \rightarrow SO(8)$.
Altogether,  the model before starting cascade breaking
has gauge group $SU(4)\times SO(8)\times (U(1)^4)$ and
has the following hypermultiplet content :  $32\, (4)+6\, (6) +123\, (1)$.
Giving generic vevs to the adjoints  and proceeding by
cascade symmetry breaking leads to the following models:
$(\hm,\vm)=\ (123,11),\ (154,10),\ (195,9),\ (272,8)$.
Again, these four models admit a $K3$ fibration interpretation in
the type II side.

{\bf  4)  $r=4$ chain}

This chain can be obtained from the $Z_6$ orbifold with a
$E_8\times E_8$ embedding given by
$V= \frac16(1,1,1,1,-4,0,0,0)\times \frac16(1,1,1,1,1,-5,0,0)$.
The resulting model has gauge group $SU(5) \times SU(4) \times U(1) \times
SU(6) \times SU(3) \times SU(2)$ and massless hypermultiplets
\beqa
\label{spz6}
\theta^0&:&(1,\ov{4},-5;1,1,1)+(10,4,1;1,1,1)+(1,1,0;6,3,2)\
+2\, (1,1,0;1,1,1)\nonumber\\
\theta^1&:& (1,1,\frac{10}3;1,3,2)+(1,4,-\frac53;6,1,1)
 + 2\, (1,1,\frac{10}3;6,1,1)  \nonumber\\
\theta^2&:& 5\, (1,\ov{4},\frac53;1,\ov{3},1)+
4\, (\ov{5},1,-\frac43;1,\ov{3},1) \nonumber\\
\theta^3&:& 3\, (1,6,0;1,1,2) + 5\, (5,1,-2;1,1,2)
\eeqa
At this point we can Higgs away most of the symmetry to arrive
at $SU(5) \times (U(1)^4)$ with essentially
$4(10) + 22(5) + 118(1)$ hypermultiplets. Giving vevs to adjoint
scalars and implementing cascade breaking leads to models with
$(\hm,\vm)=\ (118,8),\ (139,7),\ (162,6),\ (191,5),\ (244,4)$. The four
last elements fall into the $K3$ fibrations classes of ref.~\cite{klm}.

Notice that the last model in this chain is identical to the rank four
example discussed in detail in ref.~\cite{kv} . In that reference
the heterotic dual was obtained from compactification on $K3 \times T^2$
with a rank two bundle embedded in each $E_8$.
The existence of this chain suggests that this
well studied model could also be continuously
connected to the three CY compactifications
with $(b_{21},b_{11})=(190,4),(161,5)$  and $(138,6)$.
\vfill\eject

{\bf  5)  $r=11$ chain }

A simple way to construct this chain is to begin with
example 7 in ref.~\cite{kv} in which the $E_8\times E_8 $ heterotic
string is compactified  on $K3\times T^2$ and the $U(1)^4$  generic
symmetry  coming from $T^2$ is enhanced to $SU(2)\times U(1)^3$
by choosing a modulus value $T=\frac{1}{2} +i\frac{{\sqrt  3}}{2}$.
In addition,  $SU(2)$ bundles are embedded
in the first $E_8$ and in the enhanced $SU(2)$.
The gauge group at this level is $E_7\times E_8 \times U(1)^3$
with $8\, (56) + 65\, (1)$ hypermultiplets, as follows from
the index theorem.
Higgsing step by step we find the chain
$(\hm,\vm)=\ (62,18),\ (77,17),\ (90,16),\ (101,15),
\ (140,14),\ (187,13)$, $(252,12),\ (377,11)$. The last four
models again correspond to $K3$ fibrations in \cite{klm}.

\medskip

The five chains of models are collected in Table 1.
To give an idea of the starting structure,
the full gauge group before turning on generic vevs for the
Cartan subalgebra is shown for each model
as constructed above. The actual gauge group is purely Abelian.
The CY threefolds ($K3$ fibrations) that match the numbers $(\hm, \vm)$
are labelled by their weights in projective space.
When the CYs are given by a hypersurface in $\cp4 (1,1,2k_2,2k_3,2k_4)$,
the $K3$ fiber is given by a hypersurface in $\cp3 (1,k_2,k_3,k_4)$
\cite{klm}. In these cases,
we have also recorded the polynomials of the $K3$ fibers and their
quotient singularities.

The weights $(1,1,2k_2,2k_3,\cdots )$ of the CY compactifications
follow an interesting pattern that repeats in all chains.
The first element always correspond to one of the
simple complete intersection CY spaces in weighted $\cp5$
given in ref.~\cite{klm}. The remaining elements
are simple hypersurfaces in weighted $\cp4$.
The second element is obtained by just deleting
the last variable. The third member of the chain
is obtained from the second by the replacement
$(1,1,2k_2,2k_3,2k_4)\rightarrow (1,1,2k_2,2k_3, 2k_4+2k_2)$.
The fourth element is obtained  by shifting the weights
of the third element as
$(1,1,2k_2,2k_3,2k_4)\rightarrow (1,1,2k_2,2k_3+2k_2,2k_4+2k_2)$.
In the next section we will discuss how the grouping of each element
in different chains is associated to families of
quotient singularities in the $K3$ fibers.

It is natural to ask whether
there are more CYs in the lists of ref.~\cite{klm} whose weights
follow the above pattern. Indeed, there are
five additional candidate chains of just three elements whose
weights are related as those of the first three elements in the
other five chains.  The elements of these chains are displayed
in Table 2. For each chain there are two possible choices
for the  first element, the
different $b_{21}$ and last weight are indicated inside brackets.
We have identified heterotic candidate duals for some (but
not all)  of them but we will spare the reader their construction.
Finally, in addition to these 10 chains, there are pairs of models that
also seem to be connected, one of the members is a complete
intersection CY in $\cp5$ and the other is a hypersurface in $\cp4$.
These pairs include the models with $(b_{21},b_{11})=$
$(96,12),\ (131,11)$; $(76[84],10),\ (111,9)$; $(75[81],9),\ (104,8)$;
$(70[82],6),\ (101,5)$; $(69,3),\ (86,2)$.  Again, in some cases
there are two possible choices for the first element as indicated
by the different $b_{21}$ inside brackets. From
the CY models listed in the two tables of ref.~\cite{klm}, only
those with $(b_{21},b_{11})=\ (143,7),\ (68,2)$ and $(76,8)$
do not seem to fall into any of these chains.

Some comments are in order:

1) The various chains of models are not isolated from each other but
rather seem to form a {\it web of heterotic $N=2$ models}
connected by different paths involving different directions of gauge
symmetry breaking. Let us show as an example how the heterotic models
with $(\hm,\vm)=(167,15)$ and $(272,8)$ (which belong to different
chains) can be connected.  We can start with an $SO(32)$ string
compactified on any orbifold with standard embedding.
This yields generically a gauge group
$SO(28)\times SU(2)\times U(1) \times (U(1)^4)$ and
$10\, (28,2) + 66\, (1,1)$ hypermultiplets. Higgsing down to
$SO(22)$ using the hypermultiplets and then turning on vevs
to the unbroken Cartan subalgebra we recover the model
$(167, 15)$ belonging to the first chain.  If we keep on
Higgsing as much as possible we arrive at model $(272,8)$ of  the
third chain. Thus, the  heterotic $N=2$ duals form a web of theories
connected by different paths of moduli space, as expected.

2) From the above comments it is clear that a given
heterotic model can be obtained starting
from compactifications that look very different from
several points of view: different initial gauge group,
different orbifolds, etc.
Moreover, it must be pointed out that there is not a unique way to
construct a  full chain. For example, the $r=12$ chain can be
derived from any $Z_M$ orbifold with standard embedding in
$E_8 \times E_8$, or even, e.g., from $Z_3$ with a non-standard
embedding.
Also, the $r=10$ chain can be obtained by first
going to an  enhanced $SU(3)$ symmetry point and then
performing a $Z_3$ twist embedded partly in the enhanced
$SU(3)$ and partly in one $E_8$, leaving an unbroken
$E_8\times U(1)^2$, as in model $8$ of \cite{kv}.
In general there could be alternative starting models that yield the
same chains but with different (same rank) $G_r$ before adjoint Higgsing.

3) In the process of sequential gauge symmetry breaking leading to the
five chains of models in Table 1, sometimes there may be bifurcations
into other directions in Higgs space. For instance, there are examples in
which cascade symmetry breaking can proceed through an
alternative path  containing the breakings
$SU(4)\rightarrow SU(2)\times SU(2)$$\rightarrow  SU(2)\rightarrow 0$,
instead of proceeding through an $SU(3)$ intermediate step.
This is for example the case of the second chain in  Table 1.
Proceeding through $SU(2)\times SU(2)$ gives
the model $(\hm,\vm)=(144,12)$ instead of $(168,12)$.
The former corresponds to one of the models in Table 2, showing us
another example of interconnection of different chains into
a complicated web of heterotic models.

4) Not any cascade symmetry breaking of arbitrary
$N=2$ heterotic models leads to models with corresponding
type II duals in the lists of \cite{klm}.
It is not true either that any
symmetry breaking chain ending by $SU(4)\rightarrow SU(3)
\rightarrow \cdots $, is going to give rise to heterotic models with
type II duals, only some do.  For example, we could have
tried to construct the $r=4$ chain
by starting with the rank four example of \cite{kv}.
This is an $E_7\times E_7$ compactification with
hypermultiplets $4(56,1)+4(1,56)$ $+62(1,1)$. Indeed, Higgsing
completely the $E_7^2$  gives the last
element of this chain with $(\hm,\vm)=(244,4)$.
However, trying to reproduce the preceding elements
in the chain we find models $(215,5),\ (198,6)$, $(183,7)$,
none of which have type II dual candidates in the lists of
simple $K3$ fibrations.

5) There is the possibility that heterotic models, obtained
in these symmetry-breaking chains, that do {\it not} have candidate duals
in the lists of ref.~\cite{klm}
could correspond to a more general class of CY manifolds.
It is thus sensible to look for candidate duals
to the unmatched heterotic models in more general tables
of CY spaces. We have done this check for several of
the unpaired elements mentioned above and found that they
sometimes (but not always)
match CY compactifications classified in ref.~\cite{rolf}.
In these cases, unlike in the chains reported
in the present paper, the weights of the corresponding
projective spaces do not seem to follow any obvious rule.

\section{The structure of the chains of type II duals}

At the moment we do not have a satisfactory
understanding of which conditions heterotic models must
fulfill  in order to be dual to a type II compactified in one of
the $K3$ fibrations listed in ref.~\cite{klm}.
It must be emphasized that those lists
only include manifolds that are a simple generalization of
the CYs with few moduli for which
type II/heterotic duality has been tested, they are
not supossed to be exahustive compilations of $K3$ fibrations.
Still, it would be interesting to understand the origin of
all the properties of the dual chains of models described in this note.

It is natural to try to analyze our results
in terms of the underlying  6-dimensional
string-string duality  \cite{stst} between type IIA
compactifications on $K3$ and heterotic compactifications  on $T^4$.
This duality maps the cohomology of $K3$ to the Narain lattice with
signature $(20,4)$.  The idea is that if the type IIA theory is compactified
instead on a 6-dimensional Calabi-Yau which is a $K3$
fiber on $\cp1$ , the resulting
$N=2$ theory is expected to be dual to a heterotic
compactified on a variety that looks like
$T^4$  fibered over $\cp1$. This would be
 the `adiabatic ' aproximation
suggested in ref.~\cite{vw} in trying to explain the
origin of type II/heterotic  duality in four dimensions
as a fiber-wise application of string-string duality.
It turns out that there are certain
singularities in the fibration that obstruct a direct aplication
of this adiabatic argument.
Nevertheless, the authors of ref.~\cite{vw} were able to describe
qualitatively certain  features of  type II/heterotic duality
for  the rank  3 and 4 examples of ref.~\cite{kv}.
We just briefly show here how these arguments
generalize to our examples.
The basic idea is to consider the monodromy of the cohomology
of the $K3$ around the singularities in the fibration.
String-string duality suggests that the sector of the $K3$ cohomology
invariant under the monodromy is mapped to the invariant Narain lattice
in the heterotic side.

To find the form of the $K3$ fibration, we first write the simplest
transverse polynomial for the given CY in $\cp4(1,1,2k_2,2k_3,2k_4)$
and then set  $X_0=\lambda X_1$. After redefining
$X_1 \to X_1^{1/2}$ we obtain an equation of the form
\beq
F(\lambda) X_1^d + X_2^{d/k_2} + \cdots = 0
\label{k3eq}
\eeq
where $d=1+k_2+k_3+k_4$
(for Fermat-type surfaces, such as those analyzed in \cite{vw}, the mirror
CY gives rise to the same $K3$ equation).
For generic $\lambda$, this resulting equation
describes a $K3$ in weighted $\cp3$. The original CY is thus a
$K3$ fibration over the $\cp1$ parametrized by $\lambda$.
The fibration is singular when the coefficient of $X_1$ vanishes,
otherwise it can be absorbed in $X_1$. Near a simple zero we
can write $F(\lambda) = \epsilon e^{i\theta}$ with $\epsilon \to 0$.
To study the monodromy around this zero we make the transport
$\theta \to \theta + 2\pi$. Now, since $F(\lambda)$ multiplies $X_1^d$,
this is equivalent to the transformation $X_1 \to \zeta X_1$, where
$\zeta = e^{2i\pi/d}$. In the following we will just consider
$K3$ defining equations independent of $\lambda$ and analyze
the monodromy through the operation $X_1 \to \zeta X_1$.

The $K3$'s that appear in our models are
defined by transverse polynomials
in weighted $\cp3$
and have been studied in ref.~\cite{ponja}.
The 20 (1,1) forms are the K\"ahler form, forms related to polynomial
deformations of the hypersurface defining equation, and forms
related to resolution of quotient singularities.
In general, there are
$(19-s)$ polynomial deformations, where $s$ is the total rank
of the singularities due to fixed sets in the weighted
$\cp3$ that intersect the hypersurface at isolated points.
For example, in the model (492,12),
there is a singular point at $(0,0,-1,1)$ associated with
a $Z_7$ action. This singularity is of type $A_6$, it is
resolved by excising the point and glueing in 6 copies of $\cp1$
that intersect according to the Cartan matrix of $A_6$. This
example has $s=9$ since there are also
$A_1$ and $A_2$ singular points at $(0,1,-1,0)$ and $(0,-1,0,1)$,
associated with $Z_2$ and $Z_3$ actions.
As another example, consider the model (162,6).
In this case there is only an $A_3$ singularity at $(0,0,0,1)$
associated to a $Z_4$ action.

Taking into account their dependence on $X_1$,
it is straightforward to show that all polynomial deformations
transform under the monodromy. The K\"ahler form and the
$s$\, $(1,1)$ forms supported by the glued $\cp1$s,
are instead invariant. It can also be shown that the
$(0,2)$ and $(2,0)$ forms are not invariant whereas the $(0,0)$,
and $(2,2)$ forms are invariant \cite{vw}. Altogether,
there are $s+3$ invariant forms with signature $(s+1,2)$.
This is expected to be mapped to a heterotic string with an
$\Gamma(s+1,2)$ invariant lattice \cite{vw}, in agreement with
the structure
of the heterotic compactifications considered.
For instance, in the (492,12) model,
the invariant Narain lattice is $\Gamma(10,2)$, corresponding to the two
unrotated left- and right-moving coordinates plus the eight-dimensional
lattice of the unbroken gauge group.

The above discussion implies that the quotient singularities
of the $K3$ are related to vector multiplets in the
heterotic side and it can be seen as a complicated
way of computing the number of unbroken $U(1)$s as
$n_V=s+3$, or equivalently $b_{11}=s+2$.
However, there is more information regarding
the structure of the quotient singularities. In Tables 1 and 2
these singularities are given when the  CY
is a hypersurface in $\cp4$. It is evident that
to each $SU(3)\rightarrow SU(2)\rightarrow 0$
heterotic cascade, there corresponds a  `embedded'
singularity chain $A_i\rightarrow A_{i-1}\rightarrow A_{i-2}$
of the $K3$ in the type II side.
This could be related to some physical process occurring in the
type II string compactification.

\section{Final comments and conclusions}

We have identified a number of new candidates for $N\!=\!2$,
$D\!=\!4$ type II/heterotic dual pairs.
Besides the matching of the
number of hypermultiplets and vector multiplets,
we can claim
further evidence from
the fact that the heterotic duals come in symmetry-breaking chains
that are mapped into type II compactifications
on CY spaces that are $K3$ fibrations and also seem
to be related to each other in a sequential manner.

This parallelism between transitions taking place on both the type II and
the heterotic side  is reminiscent of the ideas put forward in
refs.~\cite{ast, gms}.
It is believed that the resolution of
conifold singularities in type IIB CY compactifications,
through the appearence of massless blackhole hypermultiplets, is
the dual of  the Seiberg-Witten \cite{sw}
mechanism in which massless
monopoles appear at certain strong coupling points in the
vector multiplet moduli space of the heterotic side.
In the class of models described in this note
the situation is not exactly the same.
The transitions that we have in the heterotic side are
very specific: they occur at {\it weak coupling}  and correspond to
sequential  $SU(4)\rightarrow SU(3)$ $\rightarrow SU(2)\rightarrow 0$
symmetry breaking through Higgsing of
hypermultiplet scalars. Each $SU(n)$ group is then broken by Higgsing
of vector multiplets. At points in the vector multiplet moduli space,
where the non-Abelian groups are restored, they are
very asymptotically non-free.
It would  be very interesting to understand  the dual to these
transitions in the tipe II side, probably in terms of new connections
among different CY spaces.

There is an apparent lack of uniqueness in the mapping between
heterotic and type II in the following sense.  We can start with
differently looking heterotic models before Higgsing  (i.e., different $Z_M$
orbifolds with different gauge embeddings) and  obtain after Higgsing
models that appear equally good candidates to be dual to a given
type II model on a certain CY (i.e., same number of hypermultiplets and
vector multiplets). Thus,  these various heterotic constructions seem to have
the same type II dual at strong coupling.
An example of this is the first chain in Table 1
that may be obtained equally well from any of the $Z_M$ orbifolds
with standard embedding or even from non-standard embeddings.
We could say that the dual loses
information about at least some of the
symmetries of the original heterotic model.  This is perhaps not so
surprising since in the Higgsing process we give vevs to
hypermultiplets that  carry e.g. $Z_N$ charges, so that the
original discrete symmetries of the orbifold
are expected to be spontaneously broken.

Another point to remark is that the last elements of each heterotic
chain are in some sense
more generic (in the hypermultiplet and
vector multiplet  moduli) than the preceding
elements in each chain. The gauge symmetry of these
last elements in each chain
cannot be further broken and we cannot  continue increasing the number of
massless hypermultiplets (and reducing the number of vector multiplets).
In this sense, it is amusing to note that the  heterotic $N=2$ dual with
the maximum number of massless hypermultiplets
that can be constructed is the last element in
the first chain  with $(\hm,\vm)=(492, 12)$.
This corresponds to the known CY compactification
with maximum Euler characteristic $|\chi |=960$,
that was conjectured in ref.~\cite{rolf} to be the maximum
achievable in CY compactifications.

There is also an interesting connection with cancellation
of anomalies in $N=1$ supergravity coupled to vector and
hypermultiplets in 6 dimensions. Indeed, the {\it symmetric} orbifold
heterotic models described above can be thought of as compactifications
taking place in a two step process, first to $D=6$, $N=1$ upon
compactification on $T^4/Z_M$ and then down to $D=4$, $N=2$ after
further toroidal compactification.
In $D\!=\!6$ there are strong constraints, imposed
by cancellation of gravitational anomalies, which require that
the  difference between the number of $N\!=\!1$  hypermultiplets
and vector multiplets be equal to $244$ \cite{gsw, walton, erler}
\footnote{We thank
J. Schwarz for pointing out this fact to us.}.
After compactification on a generic $T^2$,
the massless states just arrange into the appropriate
multiplets and the difference between the number of $N\!=\!2$,
$D\!=\!4$ hypermultiplets and vector multiplets, without including
the dilaton and toroidal $U(1)$s, must still be equal to $244$.
This is certainly the case for all the symmetric orbifolds
that we discussed. This argument
applies only to the original (un-Higgssed) model,
since in  $D=4$ there is not an analogous purely gravitational anomaly
constraint.  This explains why there are various  heterotic
$N=2$ models leading upon Higgsing to $(\hm,\vm)=(244,4)$.
Any model in which it is posssible to completely Higgs the
rank 16 heterotic group is bound to yield 244 hypermultiplets
due to the mentioned constraint.

Many questions concerning this class of new type II/heterotic dual pairs
still remain. In particular, it would be interesting to fully understand the
physical process ocurring in the type II side corresponding to these
heterotic chains.
One would expect that the CY manifolds in a chain are
somehow connected. Perhaps the best framework to address this
question is that of toric geometry in which a criterion for
singularity transitions can be formulated \cite{bkk, avram}.
Eventually, one would like to have
a direct method to obtain dual pairs in a systematic way.
We hope that the new examples and regularities discussed in this
note will shed some light in these issues.

\vskip0.6cm

\centerline{\bf Acknowledgments}
\bigskip

We acknowledge useful conversations with T. Banks,
M. Douglas, A. Klemm, W. Lerche, J. Schwarz and E. Witten.
We are grateful to P. Candelas and X. de la Ossa for helpful
remarks and explanations.
G.A. thanks the Departamento de F\'{\i}sica Te\'orica
at UAM for hospitality, and the Ministry of Education
and Science of Spain
as well as CONICET (Argentina) for financial support.
A.F. thanks CONICIT (Venezuela) for a research grant S1-2700.
L.E.I. would like to thank the Physics Department
of Rutgers University where part of this work was performed.

\newpage


\newpage

\begin{table}
\begin{minipage}[t]{8.0in}
\rotate[l]{
\vbox{
\begin{center}
\footnotesize
\begin{tabular}{|c|c|c|c|c|}
\hline
Group
& $(\hm, \vm) $ &  CY weights & $K3$  fiber &  $K3$ Singularity \\
\hline
\hline
$SU(4)\times E_8\times U(1)^4$ & $(167,15)$ &  $(1,1,12,16,18,20)$
  &   &
\\
\hline
$SU(3)\times E_8\times U(1)^4$ & $(230,14)$ &  $(1,1,12,16,18)$
& $X_1^{24}+ X_2^{4}+ X_3^3 + X_4^2X_2 = 0 $ & $A_1+A_2+A_8$
\\
\hline
$SU(2)\times E_8\times U(1)^4$ & $(319,13)$ &  $(1,1,12,16,30)$
& $X_1^{30}+ X_2^{5}+ X_3^3X_2 + X_4^2 = 0$ & $A_1+A_2+A_7$
\\
\hline
$ E_8\times U(1)^4$ & $(492,12)$ &  $(1,1,12,28,42)$
& $X_1^{42}+ X_2^{7}+ X_3^3+ X_4^2 = 0 $ &  $A_1+A_2+A_6$
\\
\hline
\hline
$SU(4)\times E_6\times U(1)^4$ & $(129,13)$ &  $(1,1,6,10,12,14)$
&  &
\\
\hline
$SU(3)\times E_6\times U(1)^4$ & $(168,12)$ &  $(1,1,6,10,12)$
& $X_1^{15}+ X_2^5 + X_3^3 + X_4^2X_2 = 0 $ &  $2A_2+A_5$
\\
\hline
$SU(2)\times E_6\times U(1)^4$ & $(221,11)$ &  $(1,1,6,10,18)$
& $X_1^{18}+ X_2^6 + X_3^3X_2 + X_4^2 = 0$  &  $2A_2+A_4$
\\
\hline
$ E_6\times U(1)^4$ & $(322,10)$ &  $(1,1,6,16,24)$
& $X_1^{24}+ X_2^8 + X_3^3+ X_4^2 = 0 $ &  $2A_2+A_3$
\\
\hline
\hline
$SU(4)\times SO(8)\times U(1)^4$ & $(123,11)$ &  $(1,1,4,8,10,12)$
& &
\\
\hline
$SU(3)\times SO(8)\times U(1)^4$ & $(154,10)$ &  $(1,1,4,8,10)$
& $X_1^{12}+ X_2^6 + X_3^3+ X_4^2X_2 = 0 $&  $3A_1+A_4$
\\
\hline
$SU(2)\times SO(8)\times U(1)^4$ & $(195,9)$ &  $(1,1,4,8,14)$
& $X_1^{14}+ X_2^7 + X_3^3X_2 + X_4^2 = 0 $ &  $3A_1+A_3$
\\
\hline
$ SO(8)\times U(1)^4$ & $(272,8)$ &  $(1,1,4,12,18)$
& $X_1^{18}+ X_2^9 + X_3^3+ X_4^2 = 0 $ &  $3A_1+A_2$
\\
\hline
\hline
$SU(4)\times U(1)^4$ & $(139,7)$ &  $(1,1,2,6,8,10)$
&  &
\\
\hline
$SU(3)\times U(1)^4$ & $(162,6)$ &  $(1,1,2,6,8)$
& $X_1^{9}+ X_2^9 + X_3^3+ X_4^2X_2 = 0 $ &  $A_3 $
\\
\hline
$SU(2)\times U(1)^4$ & $(191,5)$ &  $(1,1,2,6,10)$
& $X_1^{10}+ X_2^{10} + X_3^3 X_2 + X_4^2 = 0 $&   $A_2 $
\\
\hline
$  U(1)^4$ & $(244,4)$ &  $(1,1,2,8,12)$
& $X_1^{12}+ X_2^{12} + X_3^3+ X_4^2 = 0 $  &   $A_1 $
\\
\hline
\hline
$SU(4)\times E_8\times U(1)^3$ & $(140,14)$ &  $(1,1,8,12,14,16)$
  &   &
\\
\hline
$SU(3)\times E_8\times U(1)^3$ & $(187,13)$ &  $(1,1,8,12,14)$
& $X_1^{18}+ X_2^3 + X_3^3X_2 + X_4^2X_3 = 0  $ & $A_1+A_3+A_6$
\\
\hline
$SU(2)\times E_8\times U(1)^3$ & $(252,12)$ &  $(1,1,8,12,22)$
& $X_1^{22}+ X_2^2 + X_3^4X_4 + X_4^3X_3 = 0 $ & $A_1+A_3+A_5$
\\
\hline
$ E_8\times U(1)^3$ & $(377,11)$ &  $(1,1,8,20,30)$
 & $X_1^{30}+X_2^3+ X_3^5X_2 + X_4^2 = 0 $ & $A_1+A_3+A_4$
\\
\hline
\end{tabular}
\end{center}
\caption{Chains of
type IIA/heterotic
duals denoted by $ (\hm,\vm)$.
The corresponding $K3$ fibers and their singularities are also
shown for the examples which are hypersurfaces in $\cp4$.
\label{tabla1} }
}}
\end{minipage}
\end{table}

\begin{table}
\begin{center}
\begin{tabular}{|c|c|c|}
\hline
 $(b_{21},b_{11}) $ &
CY weights  & $K3$ singularity \\
\hline
\hline
 $(117[121],13)$ &  $(1,1,8,10,12,12[14])$ &
\\
\hline
 $(164,12)$ &  $(1,1,8,10,12)$ & $A_1+A_4+A_5$
\\
\hline
 $(227,11)$ &  $(1,1,8,10,20)$  &  $A_1+2A_4 $
\\
\hline
\hline
 $(104[108],12)$ &  $(1,1,6,8,10,10[12])$ &
\\
\hline
 $(143,11)$ &  $(1,1,6,8,10)$ & $A_2+A_3+A_4 $
\\
\hline
 $(194,10)$ &  $(1,1,6,8,16)$ & $A_2+2A_3 $
\\
\hline
\hline
 $(94[98],10)$ &  $(1,1,4,6,8,8[10])$  &
\\
\hline
 $(125,9)$ &  $(1,1,4,6,8)$  &  $2A_1+A_2+A_3$
\\
\hline
 $(164,8)$ &  $(1,1,4,6,12)$  &  $2A_1+2A_2$
\\
\hline
\hline
 $(98[102],6)$ &  $(1,1,2,4,6,6)$  &
\\
\hline
 $(121,5)$ &  $(1,1,2,4,6)$  &  $A_1+A_2$
\\
\hline
 $(148,4)$ &  $(1,1,2,4,8)$   &  $2A_1$
\\
\hline
\hline
 $(76[84],4)$ &  $(1,1,2,2,4,4)$  &
\\
\hline
 $(99,3)$ &  $(1,1,2,2,4)$  &  $A_1$
\\
\hline
 $(128,2)$ &  $(1,1,2,2,6)$   &  $0$
\\
\hline
\end{tabular}
\end{center}
\caption{Additional candidate chains of $K3$ fibrations denoted by
$(b_{12},b_{11})$.}
\label{tabla2}
\end{table}


\begin{thebibliography}{99}
%
\bibitem{kv}
S. Kachru and C. Vafa, hep-th/9505105.
%
\bibitem{fhsv}
S. Ferrara, J. Harvey, A. Strominger and C. Vafa, hep-th/9505162.
%
\bibitem{klm}
A. Klemm, W. Lerche, and P. Mayr,  hep-th/9506112.
%
\bibitem{klt}
V. Kaplunovsky, J. Louis and S. Theisen,  hep-th/9506110.
%
\bibitem{vw}
C. Vafa and E. Witten,  hep-th/9507050.
%
\bibitem{agnt}
I. Antoniadis, E. Gava, K.S. Narain and T.R. Taylor, hep-th/9507115.
%
\bibitem{kklmv}
S. Kachru, A. Klemm, W. Lerche, P. Mayr and C. Vafa,  hep-th/9508155;\\
I. Antoniadis and H. Partouche,  hep-th/9509009.
%
\bibitem{coni}
P. Candelas, A.M. Dale, C.A. L\"utken and R. Schimmrigk,
\NPB{298} {88} {493}; \\
P.S. Green and T. H\"ubsch, \PRL {61} {88} {1163},
Comm. Math. Phys. {\bf 119} (1988) 431; \\
P. Candelas, P.S. Green and T. H\"ubsch, \PRL {62} {89} {1956};
\NPB{330}{90}{49}.
%
\bibitem{ast}
A. Strominger,  hep-th/9504047.
%
\bibitem{gms}
B.R. Greene, D.R. Morrison and A. Strominger,  hep-th/9504145.
%
\bibitem{bbs}
K. Becker, M. Becker and A. Strominger, hep-th/9507158.
%
\bibitem{stst}
C. Hull and P. Townsend, \NPB{438} {95} {109},  hep-th/9410167;\\
E. Witten, hep-th/9503124; \\
A. Sen, hep-th/9504027;\\
J. Harvey and A. Strominger, hep-th/9504047.
%
\bibitem{schell}
B. Schellekens (ed.),{\it Superstring Construction}, North-Holland
(1989).
%
\bibitem{dhvw}
L. Dixon, J.A. Harvey, C. Vafa and E. Witten, \NPB{274} {86} {285}.
%
\bibitem{homy}
T.J. Hollowood and R.G. Myhill, Int. J. Mod. Phys. {\bf A3} (1988), 899.
%
\bibitem{kat}
Y. Katsuki, Y. Kawamura, T. Kobayashi, Y. Ono, K. Tanioka and  N. Ohtsubo
\PLB{218}{89}{169}.
%
\bibitem{sierra}
L.E. Ib\'a\~nez, J. Mas, H.P. Nilles and F. Quevedo, \NPB{301}{88}{157};\\
A. Font, L.E. Ib\'a\~nez, F. Quevedo and A. Sierra, \NPB{254}{85}{327}.
%
\bibitem{ponja}
M. Reid, Journ\'ees de Geom\'etrie Alg\'ebrique d'Angers,
(A. Beauville, ed.),  Sijthoff and Noordhoff (1980); \\
T. Yonemura, T\^ohoku. Math. J. {\bf 42} (1990) 351.
%
\bibitem{rolf}
A. Klemm and R. Schimmrigk, \NPB{411}{94}{559}, hep-th/9204060.
%
\bibitem{sw}
N. Seiberg and E. Witten, \NPB{426}{94}{19},  hep-th/9407087.
%
\bibitem{gsw}
M.B. Green, J.H. Schwarz and P.C. West, \NPB{342}{90}{246}.
%
\bibitem{walton}
M.A. Walton, \PRD{37}{87}{377}.
%
\bibitem{erler}
J. Erler, J. Math. Phys. {\bf 35} (1994), 1819, hep-th/9304104.
%
\bibitem{bkk}
P. Berglund, S. Katz and A. Klemm, hep-th/9506091.
%
\bibitem{avram}
A. Avram, P. Candelas, D. Jan\v ci\'c and M. Mandelberg, to be published.

\end{thebibliography}
\end{document}